\documentclass[journal, twoside, onecolumn, 10pt]{IEEEtran} 
\usepackage{epsf,cite,amsmath,amscd,amssymb,graphicx,latexsym,multicol}
\usepackage{algorithm,algorithmic}
\usepackage{comment}

\centerfigcaptionstrue

\begin{document}
\bibliographystyle{ieeetr}

\title{A Game-Theoretic Approach to Energy-Efficient Power Control in Multi-Carrier CDMA Systems}

\author{Farhad~Meshkati,~\IEEEmembership{Student Member,~IEEE,} Mung Chiang,~\IEEEmembership{Member,~IEEE,}
H.~Vincent~Poor,~\IEEEmembership{Fellow,~IEEE,}
and~Stuart~C.~Schwartz,~\IEEEmembership{Life Fellow,~IEEE\vspace{2cm}}
\thanks{This research was supported by the National Science Foundation under Grants ANI-03-38807,
CNS-04-27677 and CCF-04-48012. Parts of this work have been
presented at the 2005 IEEE Wireless Communication and
Networking Conference, New Orleans, LA, March 14--17, 2005.}
\thanks{The authors are with
the Department of Electrical Engineering, Princeton University,
Princeton, NJ 08544 USA (e-mail:
{\{meshkati,chiangm,poor,stuart\}@princeton.edu}).}}


\maketitle

\begin{abstract}
A game-theoretic model for studying power control in multi-carrier
CDMA systems is proposed. Power control is modeled as a
non-cooperative game in which each user decides how much power to
transmit over each carrier to maximize its own utility. The
utility function considered here measures the number of reliable
bits transmitted over all the carriers per Joule of energy
consumed and is particularly suitable for networks where energy
efficiency is important. The multi-dimensional nature of users'
strategies and the non-quasiconcavity of the utility function make
the multi-carrier problem much more challenging than the
single-carrier or throughput-based-utility case. It is shown that,
for all linear receivers including the matched filter, the
decorrelator, and the minimum-mean-square-error (MMSE) detector, a
user's utility is maximized when the user transmits only on its
``best" carrier. This is the carrier that requires the least
amount of power to achieve a particular target
signal-to-interference-plus-noise ratio (SINR) at the output of
the receiver. The existence and uniqueness of Nash equilibrium for
the proposed power control game are studied. In particular,
conditions are given that must be satisfied by the channel gains
for a Nash equilibrium to exist, and the distribution of the users
among the carriers at equilibrium is also characterized. In
addition, an iterative and distributed algorithm for reaching the
equilibrium (when it exists) is presented. It is shown that the
proposed approach results in significant improvements in the total
utility achieved at equilibrium compared to a single-carrier
system and also to a multi-carrier system in which each user
maximizes its utility over each carrier independently.
\end{abstract}


\newpage
\section{Introduction}
Power control is used for resource allocation and interference
management in both the uplink and downlink of code division multiple
access (CDMA) systems. In the uplink, the purpose of power control
is to allow each user to transmit enough power so that it can
achieve the required quality of service (QoS) at the uplink receiver
without causing unnecessary interference to other users in the
system. One of the key issues in wireless system design is energy
consumption at users' terminals. Since in many scenarios, the users'
terminals are battery-powered, efficient energy management schemes
are required in order to prolong the battery life. Hence, power
control plays an even more crucial role in such systems. Recently,
game theory has been used to study power control in data networks
and has been shown to be a very effective tool for examining this
problem (see for example \cite{MackenzieWicker01,
GoodmanMandayam00,JiHuang98,Saraydar02,
Xiao01,Zhou01,Alpcan,Sung,Meshkati_TCOMM}). In
\cite{MackenzieWicker01}, the authors provide some motivation for
using game theory to study communication systems, and in particular
power control. In \cite{GoodmanMandayam00} and \cite{JiHuang98},
power control is modeled as a non-cooperative game in which users
choose their transmit powers in order to maximize their utilities,
where utility is defined as the ratio of throughput to transmit
power. In \cite{Saraydar02}, pricing is introduced to obtain a more
efficient solution. Similar approaches are taken in
\cite{Xiao01,Zhou01,Alpcan,Sung} for different utility functions. In
\cite{Meshkati_TCOMM}, the authors extend the approach
in\cite{GoodmanMandayam00} to study the cross-layer problem of joint
multiuser detection and power control.

Multi-carrier CDMA, which combines the benefits of orthogonal
frequency division multiplexing (OFDM) with those of CDMA, is
considered to be a potential candidate for next generation high
data-rate wireless systems (see \cite{HaraPrasad97}). In particular,
in multi-carrier direct-sequence CDMA (DS-CDMA),  the data stream
for each user is divided into multiple parallel streams. Each stream
is first spread using a spreading sequence and is then transmitted
on a carrier \cite{DaSilvaSousa}. In a single user scenario with a
fixed total transmit power, the optimal power allocation strategy
for maximizing the rate is waterfilling over the frequency channels
\cite{CoverThomas}. The multiuser scenario is more complicated. In
\cite{Sriram01, Munz02, Shen03}, for example, several waterfilling
type approaches have been investigated for multiuser systems to
maximize the overall throughput. However, there are many practical
situations where enhancing power efficiency is more important than
maximizing throughput. For such applications, it is more important
to maximize the number of bits that can be transmitted per Joule of
energy consumed rather than to maximize the throughput.

Consider a multiple-access multi-carrier DS-CDMA network where
each user wishes to locally and selfishly choose its transmit
powers over the carriers in such a way as to maximize its own
utility. However, the strategy chosen by a user affects the
performance of other users in the network through multiple-access
interference. There are several questions to ask concerning this
interaction. First of all, what is a reasonable choice of a
utility function that measures energy efficiency in a
multi-carrier network? Secondly, given such a utility function,
what strategy should a user choose in order to maximize its
utility? If every user in the network selfishly and locally picks
its utility-maximizing strategy, will there be a stable state at
which no user can unilaterally improve its utility (Nash
equilibrium)? If such a state exists, will it be unique? What will
be the distribution of users among the carriers at such an
equilibrium?

Because of the competitive nature of the users' interaction, game
theory is the natural framework for modeling and studying such a
power control problem. This work is the first game-theoretic
treatment of power control in multi-carrier CDMA systems. We
propose a non-cooperative power control game in which each user
seeks to choose its transmit power over each carrier to maximize
its overall utility. The utility function here is defined as the
ratio of the user's total throughput to its total transmit power
over all the carriers. This utility function, which has units of
\emph{bits/Joule}, measures the total number of reliable bits
transmitted per Joule of energy consumed and is particularly
suitable for applications where saving power is critical. Because
of the non-cooperative nature of the proposed game, no
coordination among the users is assumed. Compared to prior work on
non-cooperative power control games, there are two difficulties to
the problem studied in this paper. One is that users' strategies
in the multi-carrier case are vectors (rather than scalars) and
this leads to an exponentially larger strategy set for each user
(i.e., many more possibilities). Secondly, the energy efficiency
utility function, which is considered here, is non-quasiconcave.
This means that many of the standard theorems from game theory as
well as convex optimization cannot be used here. In this work, we
derive the Nash equilibrium \cite{FudenbergTiroleBook91} for the
proposed power control game and study its existence and
uniqueness. We also address the following questions. If there
exists a Nash equilibrium for this game, can the users reach the
equilibrium in a distributive manner? What kind of carrier
allocations among the competing users will occur at a Nash
equilibrium? Will there be an even spread of usage of the carriers
among users? How does the performance of this joint maximization
of utility over all the carriers compare with that of an approach
where utility is maximized independently over each carrier? How
does a multi-carrier system compare with a single-carrier system
in terms of energy efficiency?

The rest of this paper is organized as follows. In
Section~\ref{background}, we provide some background for this work
by discussing the power control game for the single-carrier case.
The power control game for multi-carrier systems is presented in
Section~\ref{NCPCG}. The Nash equilibrium and its existence for
the proposed game are discussed in Sections~\ref{Nash equilibrium}
and \ref{existence NE}, respectively. In particular, in
Section~\ref{Nash equilibrium}, we derive the utility maximizing
strategy for a single user when all the other users' transmit
powers are fixed. In Section \ref{existence NE}, we show that
depending on the channel gains, the proposed power control game
may have no equilibrium, a unique equilibrium, or more than one
equilibrium, and we derive conditions that characterize the
existence and uniqueness of Nash equilibrium for a matched filter
receiver. The case of two-carrier systems is studied in more
detail in Section~\ref{special case}, where we obtain explicit
expressions for the probabilities corresponding to occurrence of
various possible Nash equilibria. In Section~\ref{algorithm}, we
present an iterative and distributed algorithm for reaching the
Nash equilibrium (when it exists). Numerical results are presented
in Section~\ref{simulation}. We show that at Nash equilibrium,
with a high probability, the users are evenly distributed among
the carriers.  We also demonstrate that our proposed method of
jointly maximizing the utility over all the carriers provides a
significant improvement in performance compared with a
single-carrier system as well the multi-carrier case in which each
user simply optimizes over each carrier independently. Finally,
conclusions are given in Section~\ref{conclusion}.

\section{Power Control Games in Single-Carrier Networks}\label{background}

\begin{figure}
\centering
\includegraphics[width=3.5in]{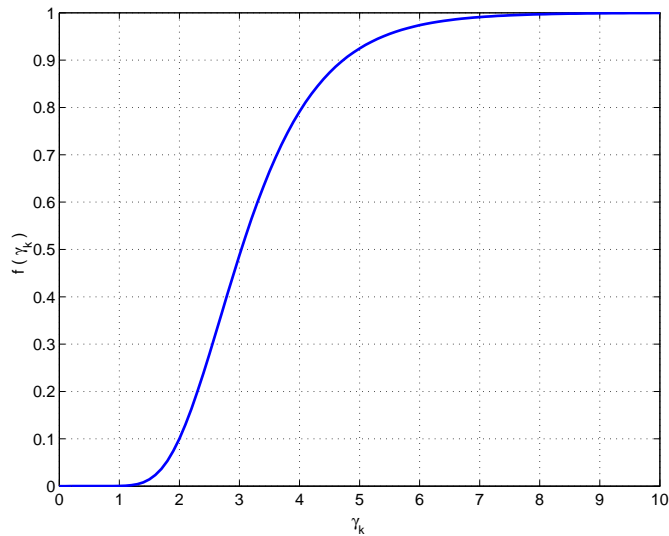}
\caption{A typical efficiency function (single-carrier case)
representing the packet success probability as a function of
received SINR.} \label{fig-eff-func}
\end{figure}

\begin{figure}
\centering
\includegraphics[width=3.5in]{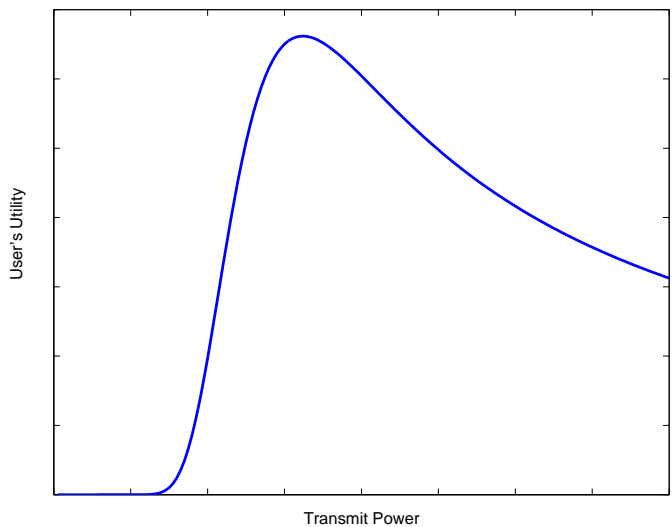}
\caption{User's utility as a function of transmit power for fixed
interference (single-carrier case).} \label{fig-util-func}
\end{figure}

Let us first look at the power control game with a single carrier.
To pose the power control problem as a non-cooperative game, we
first need to define a utility function suitable for data
applications. Most data applications are sensitive to error but
tolerant to delay. It is clear that a higher signal to interference
plus noise ratio (SINR) level at the output of the receiver will
generally result in a lower bit error rate and hence higher
throughput. However, achieving a high SINR level requires the user
terminal to transmit at a high power, which in turn results in low
battery life. This tradeoff can be quantified (as in
\cite{GoodmanMandayam00}) by defining the utility function of a user
to be the ratio of its throughput to its transmit power, i.e.,
\begin{equation}\label{eq1a}
u_k = \frac{T_k}{p_k} .
\end{equation}
Throughput is the net number of information bits that are
transmitted without error per unit time (sometimes referred to as
\emph{goodput}). It can be expressed as
\begin{equation}\label{eq1}
    T_k = \frac{L}{M} R_k f(\gamma_k)  ,
\end{equation}
where $L$ and $M$ are the number of information bits and the total
number of bits in a packet, respectively; $R_k$ and $\gamma_k$ are
the transmission rate and the SINR for the $k^{th}$ user,
respectively; and $f(\gamma_k)$ is the efficiency function
representing the packet success rate (PSR), i.e., the probability
that a packet is received without an error. Our assumption is that
if a packet has one or more bit errors, it will be retransmitted.
The efficiency function, $f(\cdot)$, is assumed to be increasing,
continuous, and S-shaped\footnote{An increasing function is S-shaped
if there is a point above which the function is concave, and below
which the function is convex.} (sigmoidal) with $f(\infty)=1$. We
also require that $f(0)=0$ to ensure that $u_k=0$ when $p_k=0$.
These assumptions are valid in many practical systems. An example of
a sigmoidal efficiency function is given in Fig. \ref{fig-eff-func}.
Using a sigmoidal efficiency function, the shape of the utility
function in \eqref{eq1a} is shown in Fig. \ref{fig-util-func} as a
function of the user's transmit power keeping other users' transmit
powers fixed. It should be noted that the throughput $T_k$ in
(\ref{eq1}) could be replaced with any increasing concave function
as long as we make sure that $u_k=0$ when $p_k=0$. A more detailed
discussion of the efficiency function can be found in
\cite{Meshkati_TCOMM}. It can be shown that for a sigmoidal
efficiency function, the utility function in (\ref{eq1a}) is a
quasiconcave\footnote{The function $f$ defined on a convex set
$\mathcal{S}$ is quasiconcave if every superlevel set of  $f$  is
convex, i.e., $\{x \in \mathcal{S} | f(x) \geq a\}$ is convex for
every value of $a$.} function of the user's transmit power
\cite{Rod03b}. This is also true if the throughput in (\ref{eq1}) is
replaced with an increasing concave function of $\gamma_k$.

Based on (\ref{eq1a}) and (\ref{eq1}), the utility function for
user $k$ can be written as
\begin{equation}\label{eq2}
    u_k = \left( \frac{L}{M} R_k \right)\frac{f(\gamma_k)}{p_k} .
\end{equation}
This utility function, which has units of \emph{bits/Joule},
captures very well the tradeoff between throughput and battery
life and is particularly suitable for applications where saving
power is crucial.

Power control is modeled as a non-cooperative game in which each
user tries to selfishly maximize its own utility. It is shown in
\cite{Saraydar02} that, in a single-carrier system, when matched
filters are used as the uplink receivers, if user terminals are
allowed to choose only their transmit powers for maximizing their
utilities, then there exists an equilibrium point at which no user
can improve its utility given the power levels of other users (Nash
equilibrium). This equilibrium is achieved when the users' transmit
powers are SINR-balanced with the output SINR being equal to
$\gamma^*$, the solution to $f(\gamma) = \gamma \ f'(\gamma)$.
Furthermore, this equilibrium is unique. In \cite{Meshkati_TCOMM},
this analysis is extended to other linear receivers.

In this work, we extend this game-theoretic approach to
multi-carrier systems. In the multi-carrier case, each user's
strategy is a vector (rather than a scalar). Furthermore, the
utility function is not a quasiconcave function of the user's
strategy. Hence, the problem is much more challenging than the one
in the single-carrier scenario.

\section{The Non-Cooperative Power Control Game in Multi-carrier Systems} \label{NCPCG}

Let us consider the uplink of a synchronous multi-carrier DS-CDMA
data network with $K$ users, $D$ carriers and processing gain $N$
(for each carrier). The carriers are assumed to be sufficiently
far apart so that the (spread-spectrum) signal transmitted over
each carrier does not interfere with the signals transmitted over
other carriers \cite{DaSilvaSousa}. We also assume that the delay
spread and Doppler spread are negligible for each individual
carrier. At the transmitter, the incoming bits for user $k$ are
divided into $D$ parallel streams and each stream is spread using
the spreading code of user $k$. The $D$ parallel streams are then
sent over the $D$ (orthogonal) carriers. For the $\ell^{th}$
carrier, the received signal at the uplink receiver (after
chip-matched filtering and sampling) can be represented by an
$N\times1$ vector as
\begin{equation}\label{eq4}
    {\mathbf{r}}_{\ell} = \sum_{k=1}^{K} \sqrt{p_{k\ell} h_{k\ell}} \ b_{k\ell}\ {\mathbf{s}}_{k} +
    {\mathbf{w}}_{\ell}
\end{equation}
where $b_{k\ell}$, $p_{k\ell}$, $h_{k\ell}$ are the $k^{th}$
user's transmitted bit, transmit power and path gain,
respectively, for the $\ell^{th}$ frequency channel (carrier);
${\mathbf{s}}_{k}$ is the spreading sequence for user $k$ which is
assumed to be random with unit norm; and ${\mathbf{w}}_{\ell}$ is
the noise vector which is assumed to be Gaussian with mean
$\mathbf{0}$ and covariance $\sigma^2 \mathbf{I}$. Let us express
the channel gain as
\begin{equation}\label{eq4b}
    h_{k \ell}= \frac{c}{d_k^\nu} a_{k \ell}^2 ,
\end{equation}
where $d_k$ is the distance of user $k$ from the uplink receiver
and $a_{k \ell}$ is a Rayleigh random variable representing the
small scale channel fading. Here, $c$ and $\nu$ are constants
which determine the path loss as a function of distance.

We propose a non-cooperative game in which each user chooses its
transmit powers over the $D$ carriers to maximize its overall
utility. In other words, each user (selfishly) decides how much
power to transmit over each frequency channel (carrier) to achieve
the highest overall utility. Let $G_D=[{\mathcal{K}}, \{A_k^{MC} \},
\{u_k^{MC} \}]$ denote the proposed non-cooperative game where
${\mathcal{K}}=\{1,\cdots, K \}$, and $A_k^{MC}=[0,P_{max}]^D$ is
the strategy set for the $k^{th}$ user. Here, $P_{max}$ is the
maximum transmit power on each carrier. Each strategy in $A_k^{MC}$
can be written as ${\mathbf{p}_k} = [p_{k1} ,\cdots, p_{kD}]$. The
utility function for user $k$ is defined as the ratio of the total
throughput to the total transmit power for the $D$ carriers, i.e.,
\begin{equation}\label{eq6}
    u_k^{MC} = \frac{\sum_{\ell=1}^D T_{k\ell}} {\sum_{\ell=1}^D p_{k\ell}} ,
\end{equation}
where $T_{k\ell}$ is the throughput achieved by user $k$ over the
$\ell^{th}$ carrier, and is given by  $T_{k\ell} = \frac{L}{M} R_k
f(\gamma_{k\ell})$ with $\gamma_{k\ell}$ denoting the received
SINR for user $k$ on carrier $\ell$. Hence, the resulting
non-cooperative game can be expressed as the following
maximization problem:
\begin{equation}\label{eq7}
    \max_{{\mathbf{p}}_k} \ u_k^{MC} =  \max_{p_{k1} ,\cdots, p_{kD} } u_k^{MC}  \ \ \ \
    \textrm{for}
   \ \  k=1,\cdots,K  ,
\end{equation}
under the constraint of non-negative powers (i.e., $p_{k\ell} \geq
0$ for all $k=1,\cdots, K$ and $\ell=1,\cdots, D$). Without
significant loss of generality, if we assume equal transmission
rates for all users, (\ref{eq7}) can be expressed as
\begin{equation}\label{eq8}
    \max_{p_{k1} ,\cdots, p_{kD} } \frac{ \sum_{\ell=1}^D f(\gamma_{k\ell})} { \sum_{\ell=1}^D p_{k\ell}} \ \ \ \
    \textrm{for}
   \ \  k=1,\cdots,K  .
\end{equation}
The relationship between the $\gamma_{k\ell}$'s and the
$p_{k\ell}$'s is
dependent on the uplink receiver. 

It should be noted that the assumption of equal transmission rates
for all users can be made less restrictive. For our analysis, it
is sufficient for the users to have equal transmission rates over
different carriers but the transmission rate can be different for
different users. More generally, the proposed power control game
can be extended to allow the users to pick not only their transmit
powers but also their transmission rates over the $D$ carriers.
While joint power and rate control is important, particularly for
data applications, our focus throughout this work is on power
control only. We will briefly comment on the joint power and rate
control problem at the end of Section~\ref{Nash equilibrium}.

\section{Nash Equilibrium for the Proposed Game}\label{Nash equilibrium}

For the non-cooperative power control game proposed in the
previous section, a Nash equilibrium is a set of power vectors,
${\mathbf{p}}_1^* , ... , {\mathbf{p}}_K^*$, such that no user can
unilaterally improve its utility by choosing a different power
vector, i.e., ${\mathbf{p}}_1^*,\cdots, {\mathbf{p}}_K^*$ is a
Nash equilibrium if and only if
\begin{equation}\label{eq9}
   u_k^{MC}({\mathbf{p}}_k^*, {{\mathbf{P}}}_{-k}^*) \geq u_k^{MC}({\mathbf{p}}_k,
    {{\mathbf{P}}}_{-k}^*) \ \textrm{for all} \ {\mathbf{p}}_k ,
\end{equation}
and for $k=1,\cdots, K$. Here, ${\mathbf{P}}_{-k}^*$ denotes the
set of transmit power vectors of all the users except for user
$k$.

We begin by characterizing utility maximization by a single user
when other users' transmit powers are fixed. \vspace{0.3cm}

\textbf{Proposition 1}: For all linear receivers and with all
other users' transmit powers being fixed, user $k$'s utility
function, given by (\ref{eq6}), is maximized when
\begin{equation}\label{eq11a}
p_{k\ell}=\left\{%
\begin{array}{ll}
    p_{k L_k}^* &  \textrm{for} \ \ \ \ell=L_k\\
    0     &  \textrm{for} \ \ \ \ell \neq L_k \\
\end{array}%
\right. ,
\end{equation}
where $L_k=\arg \min_{\ell} p_{k \ell}^*$ with $p_{k \ell}^*$
being the transmit power required by user $k$ to achieve an output
SINR equal to $\gamma^*$ on the $\ell^{th}$ carrier, or $P_{max}$
if $\gamma^*$ cannot be achieved. Here, $\gamma^*$ is the unique
(positive) solution of $f(\gamma) = \gamma \
f'(\gamma)$.\vspace{0.2cm}

\emph{Proof}: We first show that $\frac{f(\gamma)}{p}$ is
maximized when $p$ is such that $\gamma=\gamma^*$. For this, we
take the derivative of $\frac{f(\gamma)}{p}$ with respect to $p$
and equate it to zero to obtain
\begin{equation}\label{eq11}
    p  \frac{\partial \gamma}{\partial p}  f'(\gamma) - f(\gamma)
    =0 .
\end{equation}
Since for all linear receivers $p  \frac{\partial \gamma}{\partial
p} =\gamma$ \cite{Meshkati_TCOMM}, $\frac{f(\gamma)}{p}$ is
maximized when $\gamma=\gamma^*$, the (positive) solution to
$f(\gamma)=\gamma f'(\gamma)$. It is shown in \cite{Rod03b} that
for an S-shaped function, $\gamma^*$ exists and is unique. If
$\gamma^*$ cannot be achieved, $\frac{f(\gamma)}{p}$ is maximized
when $p=P_{max}$. Now, define $p_{k \ell}^*$ as the transmit power
required by user $k$ to achieve an output SINR equal to $\gamma^*$
on the $\ell^{th}$ carrier (or $P_{max}$ if $\gamma^*$ is not
achievable) and let $L_k=\arg \min_{\ell} p_{k \ell}^*$. In case
of ties, we can pick any of the indices corresponding to the
minimum power. Then, based on the above argument we have
$\frac{f(\gamma_{k L_k})}{p_{k L_k}} \leq \frac{f(\gamma^*)}{p_{k
L_k}^*}$ for any $p_{k L_k} \geq 0$. Also, because $p_{k
L_k}^*=\min_{\ell} p_{k \ell}^*$, we have $\frac{f(\gamma_{k
\ell})}{p_{k \ell}} \leq \frac{f(\gamma^*)}{p_{k \ell}^*} \leq
\frac{f(\gamma^*)}{p_{k L_k}^*}$ for all $p_{k \ell} \geq 0$ and
$\ell \neq L_k$. Based on the above inequalities, we can write
\begin{equation}\label{eq12}
\frac{f(\gamma_{k \ell})} {f(\gamma^*)}
 \leq \frac{p_{k \ell}}{p_{k L_k}^*} , \ \ \ \ \  \textrm{for} \ \ \ell=1, 2,\cdots, D .
\end{equation}
Adding the $D$ inequalities given in (\ref{eq12}) and rewriting
the resulting inequality, we have
\begin{equation}\label{eq13}
 \frac{ \sum_{\ell=1}^D f(\gamma_{k \ell})} { \sum_{\ell=1}^D p_{k \ell}}
\leq \frac{ f( \gamma^*)} {p_{k L_k}^*}  \ \ \textrm{for all} \ \
p_{k1},\cdots, p_{kD} \geq 0 \ .
\end{equation}
This completes the proof. \ ${\Box}$ \vspace{0.3cm}

Proposition 1 suggests that the utility for user $k$ is maximized
when the user transmits only over its ``best" carrier such that
the achieved SINR at the output of the uplink receiver is equal to
$\gamma^*$. The ``best" carrier is the one  that requires the
least amount of transmit power to achieve $\gamma^*$ at the output
of the uplink receiver. Based on Proposition 1, at a Nash
equilibrium each user transmits only on one carrier. This
significantly reduces the number of cases that need to be
considered as possible candidates for a Nash equilibrium. A set of
power vectors, ${\mathbf{p}}_1^* ,\cdots, {\mathbf{p}}_K^*$, is a
Nash equilibrium if and only if they simultaneously satisfy
(\ref{eq11a}).

It should also be noted that the utility-maximizing strategy
suggested by Proposition 1 is different from the waterfilling
approach that is discussed in \cite{WeiYu02} for digital subscriber
line (DSL). This is because in \cite{WeiYu02}, utility is defined as
the user's throughput and the goal there is to maximize this utility
function for a fixed amount of available power. Here, on the other
hand, the amount of available power is not fixed. In addition,
utility is defined here as the number of bits transmitted per Joule
of energy consumed which is particularly suitable for systems with
energy constraints.

Alternatively, user $k$'s utility function can be defined as
$\tilde{u}_k= \sum_{\ell=1}^D \frac{T_{k\ell}}{p_{k\ell}}$. This
utility function is maximized when each of the terms in the
summation is maximized. This happens when the user transmits on all
the carriers at power levels that achieve $\gamma^*$ for every
carrier. This is equivalent to the case in which each user maximizes
its utility over each carrier independently. We show in Section
\ref{simulation} that our proposed joint maximization approach,
through performing a distributed interference avoidance mechanism,
significantly outperforms the approach of individual utility
maximization. Throughout this paper, the expression in \eqref{eq6}
is used for the user's utility function.

Since at Nash equilibrium (when it exists), each user must
transmit on one carrier only, there are exactly $D^K$
possibilities for an equilibrium. For example, in the case of
$K=D=2$, there are four possibilities for Nash equilibrium:
\begin{itemize}
    \item User 1 and user~2 both transmit on the first carrier.
    \item User 1 and user~2 both transmit on the second carrier.
    \item User 1 transmits on the first carrier and user~2 transmits on the second carrier.
    \item User 1 transmits on the second carrier and user~2 transmits on the first carrier.
\end{itemize}

Depending on the set of channel gains, i.e., the $h_{k\ell}$'s,
the proposed power control game may have no equilibrium, a unique
equilibrium, or more than one equilibrium. In the following, we
investigate the existence and uniqueness of Nash equilibrium for
the conventional matched filter receiver and also comment on the
extensions of the results to other linear multiuser receivers such
as the decorrelating and minimum mean square error (MMSE)
detectors \cite{LupasVerdu89} \cite{MadhowHonig94}.

For the joint power and rate control problem, it can be shown by
using a similar technique as the one used in the proof of
Proposition 1 that for each user to maximize its own utility, the
user must transmit only on its ``best" carrier. Furthermore, the
combined choice of power and rate has to be such that the output
SINR is equal to $\gamma^*$. This implies that there are infinite
combinations of power and rate that maximize the user's utility
given that the powers and rates of other users are fixed.

\section{Existence and Uniqueness of Nash
Equilibrium}\label{existence NE}

If we assume random spreading sequences, the output SINR for the
$\ell^{th}$ carrier of the $k^{th}$ user with a matched filter
receiver is given by
\begin{equation}\label{eq5}
\gamma_{k\ell} = \frac{p_{k\ell} h_{k\ell} } {\sigma^2 +
\frac{1}{N}\sum_{j\neq k} p_{j\ell} h_{j\ell} }\ .
\end{equation}
Let us define
\begin{equation}\label{eq10}
    \hat{h}_{k\ell} =\frac{ h_{k\ell} } {\sigma^2 +
\frac{1}{N}\sum_{j\neq k} p_{j\ell} h_{j\ell} }
\end{equation}
as the ``effective channel gain" for user $k$ over the $\ell^{th}$
carrier. Based on (\ref{eq5}) and (\ref{eq10}), we have $\gamma_{k
\ell}= \hat{h}_{k \ell} p_{k \ell}$.

Let us for now assume that the processing gain is sufficiently
large so that even when all $K$ users transmit on the same
carrier, $\gamma^*$ can be achieved by all users. This is the case
when $N>(K-1)\gamma^*$. We later relax this assumption. The
following proposition helps identify the Nash equilibrium (when it
exists) for a given set of channel gains. \vspace{0.3cm}

\textbf{Proposition 2}: For a matched filter receiver, a necessary
condition for user $k$ to transmit on the $\ell^{th}$ carrier at
equilibrium is that
\begin{equation}\label{eq14}
    \frac{ h_{k \ell}} {h_{ki}} > \frac{\Theta_{n(\ell)} } {
    \Theta_{n(i)} } \ \Theta_0  \ \ \ \ \  \textrm{for all}\ i\neq
    \ell \ ,
\end{equation}
where $n(i)$ is the number of users transmitting on the $i^{th}$
carrier and
\begin{equation}\label{eq15}
    \Theta_n = \frac{1} { 1 - (n-1) \frac{\gamma^*}{N}} \ \ \ \ n=0,
    1,\cdots, K .
\end{equation}
In this case, $p_{k\ell}^*= \frac{\gamma^* \sigma^2}{h_{k\ell}}
 \Theta_{n(\ell)}$.\vspace{0.2cm}

\emph{Proof}: Based on Proposition 1, in order for user $k$ to
transmit on carrier $\ell$ at equilibrium, we must have
\begin{equation}\label{eq15a}
\hat{h}_{k \ell}
> \hat{h}_{k i} \ \textrm{for all} \ i \neq \ell .
\end{equation}
Since $n(\ell)$ users (including user $k$) are transmitting on the
$\ell^{th}$ carrier and $n(i)$ users are transmitting on the
$i^{th}$ carrier and all users have an output SINR equal to
$\gamma^*$, we have
\begin{equation}\label{eq15b}
\hat{h}_{k\ell} =\frac{ h_{k\ell} } {\sigma^2 + \frac{
n(\ell)-1}{N} q_{\ell} } ,
\end{equation}
and
\begin{equation}\label{eq15c}
\hat{h}_{ki} =\frac{ h_{ki} } {\sigma^2 + \frac{n(i)}{N} q_i } ,
\end{equation}
where $q_{\ell}=\frac{\sigma^2 \gamma^*}{1- \frac{( n(\ell)-1)
\gamma^*}{N}}$ and
 $q_i=\frac{\sigma^2 \gamma^*}{1- \frac{( n(i)-1)
\gamma^*}{N}}$ are the received powers for each user on the
$\ell^{th}$ and $i^{th}$ carriers, respectively. Now define
$\Theta_n= \frac{1}{1-(n-1)\frac{\gamma^*}{N}}$ to get $q_{\ell}=
\sigma^2 \gamma^* \Theta_{n(\ell)}$ and $q_i=\sigma^2 \gamma^*
\Theta_{n(i)}$. Substituting $q_{\ell}$ and $q_i$ into
(\ref{eq15b}) and (\ref{eq15c}) and taking advantage of the fact
that $1+\frac{(n-1)\gamma^*}{N}\Theta_n = \Theta_n$, we get
\begin{equation}\label{eq15d}
\hat{h}_{k\ell} =\frac{ h_{k\ell} } {\Theta_{n(\ell)} \sigma^2 } ,
\end{equation}
and
\begin{equation}\label{eq15e}
\hat{h}_{ki} =\frac{ h_{ki} } { \frac{\Theta_{n(i)}}{\Theta_0}
\sigma^2} .
\end{equation}
Consequently, (\ref{eq14}) is obtained by substituting
(\ref{eq15d}) and (\ref{eq15e}) into (\ref{eq15a}). Furthermore,
since $p_{k\ell}^* h_{k\ell} = q_{\ell} =\sigma^2 \gamma^*
\Theta_{n(\ell)}$, we have  $p_{k\ell}^*= \frac{\gamma^*
\sigma^2}{h_{k\ell}} \Theta_{n(\ell)}$, and this completes the
proof. \ ${\Box}$ \vspace{0.3cm}

Note that, based on (\ref{eq15}), when $N > (K-1) \gamma^*$, we
have ${0< \Theta_0 < \Theta_1 < \Theta_2 < \cdots < \Theta_K}$
with $\Theta_1=1$.

For each of the $D^K$ possible equilibria, the channel gains for
each user must satisfy $D-1$ inequalities similar to (\ref{eq14}).
Furthermore, satisfying a set of $K(D-1)$ of such inequalities by
the $K$ users is sufficient for existence of Nash equilibrium but
the uniqueness is not guaranteed. For example, for the case of
$K=D=2$, the four possible equilibria can be characterized as
follows.
\begin{itemize}
    \item For both users to transmit on the first carrier at equilibrium, we must have
      $\frac{h_{11}}{h_{12}}>\Theta_2$ and $\frac{h_{21}}{h_{22}}>\Theta_2$.
    \item For both users to transmit on the second carrier at equilibrium, we must have
      $\frac{h_{11}}{h_{12}}<\frac{1}{\Theta_2}$ and $\frac{h_{21}}{h_{22}}<\frac{1}{\Theta_2}$.
    \item For user 1 and user~2 to transmit on the first and second carriers, respectively, at equilibrium, we must have
     ${\frac{h_{11}}{h_{12}}>\Theta_0}$ and $\frac{h_{21}}{h_{22}}<\frac{1}{\Theta_0}$.
    \item For user 1 and user~2 to transmit on the second and first carriers, respectively, at equilibrium, we must have
      ${\frac{h_{11}}{h_{12}}<\frac{1}{\Theta_0}}$ and $\frac{h_{21}}{h_{22}}>\Theta_0$.
\end{itemize}

\begin{figure}
\centering
\includegraphics[width=3.5in]{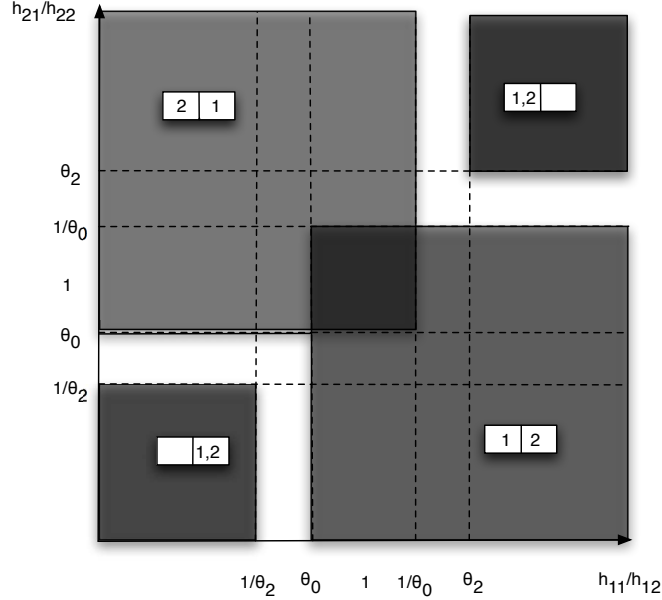}
\caption{Nash equilibrium regions for the case of two users and two
carriers. (1,2), for example, means that at equilibrium user 1
transmits on the first carrier and user~2 transmits on the second
carrier. } \label{fig2}
\end{figure}

Fig. \ref{fig2} shows the regions corresponding to the above four
equilibria. It can be seen that for certain values of channel
gains there is no Nash equilibrium (the white areas in the figure)
and for some values of channel gains there are two possible
equilibria. When the channel gains belong to the white regions,
none of the four possible candidates for Nash equilibrium is
stable. It can be shown that if we put the two users in any of
these four possible states, at least one of the users would prefer
to jump to the other carrier because that would improve its
utility. Hence, no Nash equilibrium exists.

It is interesting to observe that homogeneity of channel gains can
prevent Nash equilibrium from existing. Consider the two-user
two-carrier case where $h_{12}=h_{21}$, it is easy to verify the
following:\\ \textbf{Corollary 1}: If either $h_{11}/h_{22}$ or
$h_{22}/h_{11}$ belongs to $[1/\Theta_{2}^{2}, \Theta_{0}^{2}]$,
then there does not exist a Nash equilibrium.\vspace{0.2cm}

When the processing gain is less than or equal to $(K-1)\gamma^*$,
it is not possible for all users to transmit on the same carrier and
achieve $\gamma^*$ simultaneously. Instead, they would end up
transmitting at the maximum power (see Proposition~1). More
specifically, for any given $N\leq (K-1)\gamma^*$, at most $\lfloor
N/\gamma^*-1\rfloor$ users can simultaneously achieve $\gamma^*$ on
the same carrier. Therefore, as $N$ decreases, those Nash equilibria
in which the distribution of users among the carriers is less
uniform are less likely to occur. This means that for small values
of processing gain, the distribution of users among the carriers
becomes more uniform. However, the probability that no Nash
equilibrium exists increases. We will discuss this further in
Section~\ref{special case}.

Similar analysis can be done for the decorrelating and MMSE
detectors. However, because of the dependence of the SINR
expressions of these receivers on the spreading sequences of all the
users, obtaining closed form expressions such as the ones given in
\eqref{eq14} and \eqref{eq15} is not possible.

\section{Special Case of Two Carriers}\label{special case}

To gain some insight into the properties of the Nash equilibria
for our proposed game, let us concentrate on a system with two
carriers and two users (i.e., $D=K=2$).  We assume that the $a_{k
\ell}$'s in \eqref{eq4b} are independent and identically
distributed (i.i.d.) among the users and carriers and have the
Rayleigh distribution with mean 1. As a result, the
$a_{k\ell}^2$'s will be i.i.d. with the exponential distribution
of mean 1. Let $X_1$ be the random variable corresponding to the
number of users that transmit over the first carrier at
equilibrium. Also let $P_{X_1}(m)$ represent the probability that
there are $m$ users on the first carrier at Nash equilibrium.

\textbf{Proposition 3}: If $N>\gamma^*$, then the probabilities
that there are zero, one, and two users transmitting on the first
carrier at Nash equilibrium are, respectively, given by
\begin{eqnarray}
    P_{X_1}(0)&=& \left(\frac{1}{1+\Theta_2}\right)^2 \label{eq16} \\
    P_{X_1}(1)&=& 2\left(\frac{1}{1+\Theta_0}\right)^2-\left(\frac{1-\Theta_0}{1+\Theta_0}\right)^2 \label{eq16b}\\
    P_{X_1}(2)&=& \left(\frac{1}{1+\Theta_2}\right)^2 \ .\label{eq16c}
\end{eqnarray}

\emph{Proof}: If we assume $N>\gamma^*$, then, based on
Proposition 2, the probability that both users transmit on the
first carrier at equilibrium (i.e., $X_1=2$) is given by (see Fig.
\ref{fig2})
\begin{eqnarray}
    P_{X_1}(2)&=&\textrm{Pr}\{X_1=2\} = \textrm{Pr}\left\{\frac{h_{11}}{h_{12}}>\Theta_2,
    \frac{h_{21}}{h_{22}}>\Theta_2\right\}\nonumber\\
    &=& \textrm{Pr}\left\{\frac{a_{11}^2}{a_{12}^2}>\Theta_2\right\}
    \textrm{Pr}\left\{\frac{a_{21}^2}{a_{22}^2}>\Theta_2\right\}= \left( \underset{\{(u,v): u\geq 0, v\geq 0\ \textrm{and}\ \frac{u}{v}>\Theta_2\}}{\int \int} e^{-(u+v)} du\ dv \right)^2 \nonumber \\
    &=&
    \left(\frac{1}{1+\Theta_2}\right)^2.\hspace{0.5cm}\nonumber
\end{eqnarray}
Similarly, the probability of both users transmitting on the
second carrier at equilibrium is
\begin{eqnarray}
    P_{X_1}(0)&=&\textrm{Pr}\{X_1=0\} 
    = \textrm{Pr}\left\{\frac{h_{12}}{h_{11}}>\Theta_2, \frac{h_{22}}{h_{21}}>\Theta_2\right\}=
    \left(\frac{1}{1+\Theta_2}\right)^2.\hspace{0.5cm} \nonumber
\end{eqnarray}
The probability of one user transmitting on each of the two
carriers can be found to be
\begin{eqnarray}
    P_{X_1}(1)=\textrm{Pr}\{X_1=1\} &=& \textrm{Pr}\left\{\frac{h_{11}}{h_{12}}>\Theta_0,
    \frac{h_{21}}{h_{22}}<\frac{1}{\Theta_0}\right\}
     + \textrm{Pr}\left\{\frac{h_{11}}{h_{12}}<\frac{1}{\Theta_0},
     \frac{h_{21}}{h_{22}}>{\Theta_0}\right\}\nonumber\\
    && - \textrm{Pr}\left\{\Theta_0<\frac{h_{11}}{h_{12}}<\frac{1}{\Theta_0}, \Theta_0<\frac{h_{21}}{h_{22}}<\frac{1}{\Theta_0}\right\}\nonumber\\
    &=& 2\left(\frac{1}{1+\Theta_0}\right)^2-\left(\frac{1-\Theta_0}{1+\Theta_0}\right)^2
    .\nonumber
\end{eqnarray}
This completes the proof. ${\Box}$ \vspace{0.3cm}

Based on \eqref{eq16}--\eqref{eq16c}, the probability that no Nash
equilibrium exists is given by
\begin{eqnarray}\label{eq16d}
    P_o&=&\textrm{Pr}\{\textrm{No\ Nash\ equilibrium}\} = 2\left\{\left(\frac{\Theta_0}{1+\Theta_0}\right)^2 -
    \left(\frac{1}{1+\Theta_2}\right)^2\right\} .
\end{eqnarray}

It should be noted that since $P_o>0$, the $P_{X_1}(m)$'s, in
general, do not add up to 1. Therefore, they represent a
\emph{pseudo} probability mass function (pseudo PMF) for $X_1$. As
the processing gain $N$ becomes larger, $\Theta_0$ and $\Theta_2$
approach 1 from below and above, respectively. This results in a
reduction in $P_{X_1}(1)$ but an increase in $P_{X_1}(0)$ and
$P_{X_1}(2)$, i.e., the pseudo PMF for $X_1$ becomes flatter.
However, the increase outweighs the decrease and as a result $P_o$
decreases as $N$ increases. Going back to Fig. \ref{fig2}, we see
that the region for which no Nash equilibrium exists shrinks as $N$
increases. In addition, the region for which more than one
equilibrium exists disappears as $N$ becomes very large. Therefore,
as the processing gain becomes large, the probability that the
proposed power control game has a unique Nash equilibrium approaches
one. Note that as $N\rightarrow\infty$, we have $\Theta_0\rightarrow
1$ and $\Theta_2\rightarrow 1$. Therefore, based on
\eqref{eq16}--\eqref{eq16d}, in the limit of large processing gains,
we have $P_{X_1}(0)= P_{X_1}(2)=\frac{1}{4}$,
$P_{X_1}(1)=\frac{1}{2}$ and $P_o=0$. This is because for a
sufficiently large processing gain, the multiple-access interference
becomes insignificant compared to the background noise. Therefore,
each user transmits on the carrier with the best channel gain. This
means that the probability that user 1 transmits on the first
carrier becomes independent of whether the other user is
transmitting on the first or second carrier (and vice versa).

So far, the assumption has been that $N>\gamma^*$ so that both
users can achieve $\gamma^*$ even when they are transmitting over
the same carrier.

\textbf{Proposition 4}: If $N \leq \gamma^*$, then the
probabilities that there are zero, one, and two users transmitting
on the first carrier at Nash equilibrium are, respectively, given
by
\begin{eqnarray}
    P_{X_1}(0)&\approx& 0\label{eq16d} \\
    P_{X_1}(1)&=& 2\left(\frac{1}{1+\Theta_0}\right)^2-\left(\frac{1-\Theta_0}{1+\Theta_0}\right)^2 \label{eq16e}\\
    P_{X_1}(2)&\approx& 0 \ .\label{eq16f}
\end{eqnarray}

\emph{Proof}: If $N \leq \gamma^*$, the users cannot achieve
$\gamma^*$ simultaneously when they are transmitting on the same
carrier and they would end up transmitting at the maximum power.
Hence,
\begin{equation}
    P_{X_1}(2)=\textrm{Pr}\{X_1=2\} = \textrm{Pr}\{ P_{max}
< p_{12}^*  \ \textrm{and} \ P_{max}< p_{22}^* \} , \nonumber
\end{equation}
where $p_{12}^* =  \frac{\sigma^2 \gamma^*}{h_{12}}$ and $p_{22}^*
= \frac{\sigma^2 \gamma^*}{h_{22}}$. Therefore,
\begin{eqnarray}
   P_{X_1}(2) &=& \textrm{Pr}\left\{h_{12}  <  \frac{\sigma^2\gamma^*
   }{P_{max}} \ \textrm{and} \ h_{22}  <  \frac{\sigma^2\gamma^*
   }{P_{max}}\right\} \nonumber \\
   &=& \textrm{Pr}\left\{h_{12}  <  \frac{\sigma^2\gamma^*
   }{P_{max}}\right\}\textrm{Pr}\left\{h_{122}  <  \frac{\sigma^2\gamma^*
   }{P_{max}}\right\} \nonumber ,
\end{eqnarray}
assuming independent channel gains.

Now based on \eqref{eq4b}, we have
\begin{equation}
    P_{X_1}(2)=\textrm{Pr}\left\{a_{12}^2 < \frac{d_{1}^\nu \sigma^2
    \gamma^*}{cP_{max}} \right\} \textrm{Pr}\left\{a_{22}^2 < \frac{d_{2}^\nu \sigma^2
    \gamma^*}{cP_{max}} \right\} \nonumber
\end{equation}

If the channel amplitudes have a Rayleigh distribution, then
$a_{k\ell}^2$'s have an exponential distribution with mean 1.
Therefore, $$\textrm{Pr}\left\{a_{k\ell}^2 < \frac{d_{k}^\nu
\sigma^2 \gamma^*}{cP_{max}} \right\}=1-e^{-b_{k\ell}}$$ where
$b_{k\ell}= \frac{d_k^\nu \sigma^2 \gamma^*}{cP_{max}}$.

For typical values of $c$, $d$, $\nu$, $\sigma^2$ and $P_{max}$, $b$
is very small. For example, for $c=0.1$, $d=100$m , $\nu=4$,
$\sigma^2= 10^{-16}$W , $\gamma^*=6.4$,  and $P_{max} =1$W, we get
$b=6.4 \times 10^{-7}$. As a result, $1-e^{-b}\simeq b$. Therefore,
$P_{X_1}(2) \cong b_{12}b_{22}\approx 0$. Similarly, we have
$P_{X_1}(0) \cong b_{11}b_{21}\approx 0$.

To obtain $P_{X_1}(1)$, we can follow the same steps as the ones
used for the case of $N>\gamma^*$. Hence, when $N\leq\gamma^*$, we
have
${P_{X_1}(1)=2\left(\frac{1}{1+\Theta_0}\right)^2-\left(\frac{1-\Theta_0}{1+\Theta_0}\right)^2}$.\\This
complete the proof. ${\Box}$ \vspace{0.3cm}

Based on \eqref{eq16d}--\eqref{eq16f}, we have
$P_o=1-\left[2\left(\frac{1}{1+\Theta_0}\right)^2-\left(\frac{1-\Theta_0}{1+\Theta_0}\right)^2\right]
=2\left(\frac{\Theta_0}{1+\Theta_0}\right)^2$ when
$N\leq\gamma^*$.

We can summarize Propositions~3 and 4 as
\begin{equation}\label{eq17a}
    P_{X_1}(0)=P_{X_1}(2)= \left\{%
\begin{array}{ll}
     \  0  & \textrm{if} \ \ N\leq\gamma^* \\
    \left(\frac{1}{1+\Theta_2}\right)^2 & \textrm{if} \ \ N> \gamma^* \\
\end{array}%
\right. ,
\end{equation}
\begin{equation}\label{eq17b}
    P_{X_1}(1)=2\left(\frac{1}{1+\Theta_0}\right)^2-\left(\frac{1-\Theta_0}{1+\Theta_0}\right)^2
    ,
\end{equation}
and
\begin{equation}\label{eq17c}
    P_o=\left\{%
\begin{array}{ll}
    2\left(\frac{\Theta_0}{1+\Theta_0}\right)^2 & \textrm{if} \ \ N\leq\gamma^* \\
    2\left[\left(\frac{\Theta_0}{1+\Theta_0}\right)^2 -
    \left(\frac{1}{1+\Theta_2}\right)^2\right] & \textrm{if} \ \ N>\gamma^* \\
\end{array}%
\right.\ .\vspace{5mm}
\end{equation}

Although obtaining explicit expressions for the probabilities of
the occurrence of various Nash equilibria for the case in which
$K>2$ is more complicated, many of the general trends observed for
the $K=2$ case are also valid when $K>2$. Namely, as $N$ increases
the pseudo PMF of $X_1$ becomes wider (i.e., it has larger
variance) and at the same time the probability that no equilibrium
exists becomes smaller. This means that in the asymptotic case of
large processing gains, the proposed power control game has a
unique equilibrium. Furthermore, for very large values of $N$, the
PMF of $X_1$ can be approximated as
\begin{equation}\label{eq18}
P_{X_1}(m) = \textrm{Pr}\{X_1=m\}\approx\left(%
\begin{array}{c}
  K \\
  m \\
\end{array}%
\right)\left(\frac{1}{2}\right)^K \textrm{for} \ m=0,\cdots, K.
\end{equation}

Similar arguments can be made for the decorrelating and MMSE
detectors. In particular, since these receivers are more powerful
in combating interference, for the same processing gain, the
pseudo PMF of $X_1$ is wider as compared to the one for the
matched filter. Also, since multiuser receivers suppress
multiple-access interference more effectively, the probability of
presence of a particular user on a certain carrier is almost
independent of where the other users are transmitting. This is
particularly true for large processing gains. Therefore, the
approximation in \eqref{eq18} is more accurate for the
decorrelating and MMSE detectors as compared to the conventional
matched filter receiver.

The validity of these claims will also be confirmed through
simulations in Section \ref{simulation}.

\section{A Distributed Power Control Algorithm}\label{algorithm}

In this section, we present an iterative and distributed algorithm
for reaching a Nash equilibrium of the proposed power control game
(when it exists). This algorithm is applicable to all linear
receivers including the matched filter, decorrelating and MMSE
detectors. The description of the algorithm is as follows.

{\bf The Best-response Multi-carrier Power-control (BMP)
Algorithm}: Consider a network with $K$ users and $D$ carriers.
\begin{enumerate}
    \item Initialize the transmit powers of all the users over all
    the carriers, and let $NumIter=0$.
    \item Set $k=1$, and $NumIter=NumIter+1$.\label{step1}
    \item Given the transmit powers of other users, user $k$ picks
    its ``best" carrier and transmits only on this carrier at a power
    level that achieves an output SINR equal to $\gamma^*$. The
    ``best" carrier is the one that requires the least amount of
    transmit power for achieving $\gamma^*$.
    \label{step2}
    \item $k=k+1$.
    \item If $k \leq K$ then go back to {step~\ref{step2}}.
    \item Stop if the powers have converged or if $NumIter > MaxNumIter$; otherwise go to
    {step~\ref{step1}}.
\end{enumerate}

This is a best-response algorithm since at each stage, a user
decides to transmit on the carrier that maximizes the user's
utility (i.e., its best-response strategy) given the current
conditions of the system. In Step~\ref{step2}, it may appear that
each user needs to know not only its own transmit powers and
channel gains but also those for all the other users in order to
determine its ``best" carrier. However, it should be noted that it
is actually sufficient for the user to only know its own received
SINRs on each carrier. This information can for example be fed
back to the user terminal from the access point.

It is clear that if the above algorithm converges, it will
converge to a Nash equilibrium. The question that remains to be
answered is whether or not the above algorithm converges whenever
a Nash equilibrium exists. In Appendix~\ref{appendixI}, we have
proved that for the two-user two-carrier case, the BMP algorithm
converges to a Nash equilibrium when it exists. Using a similar
technique, we have also proved the convergence for the case of two
users and $D$ carriers as well as for the three-user two-carrier
case (see Appendices~\ref{appendixII}~and~\ref{appendixIII}). We
have shown that for each of the possible Nash equilibria, starting
from any state, the algorithm will eventually reach the
equilibrium state. Since the general case of $K$ users and $D$
carriers is not fundamentally different from the ones considered
here, the same technique can be used to prove the convergence of
the BMP algorithm for the general case of $K$ users and $D$
carriers except that there are many more possibilities to consider
in the proof. The convergence of the BMP algorithm has also been
demonstrated in Section \ref{simulation} using extensive
simulations. In the case of multiple Nash equilibria, the
algorithm converges to one of the equilibria depending on the
starting point. For the scenarios where no Nash equilibrium
exists, users keep jumping from one carrier to another one when
the BMP algorithm is used. However, the algorithm stops when the
number of iterations reaches \emph{MaxNumIter}. The value of
\emph{MaxNumIter} should be chosen large enough to ensure the
convergence of the powers for the cases where an equilibrium
exists.

\section{Simulation Results} \label{simulation}

We first consider the case of two carriers with two users. We assume
$L=M=100$, $R=100$Kbps and $\sigma^2=5\times 10^{-16}$Watts; and use
$f(\gamma)=(1-e^{-\gamma})^M$ as the efficiency
function\footnote{This is a useful example for the efficiency
function and serves as an approximation to the packet success rate
that is very reasonable for almost all practical cases with moderate
to large values of $M$.}. For this efficiency function,
$\gamma^*=6.4$ (=8.1dB). We assume that the channel gains are i.i.d.
with exponential distribution of mean 1. We consider 20 000
realizations of the channel gains. For each realization, we run the
BMP algorithm, proposed in {Section~\ref{algorithm}}, for 20
iterations (i.e., \emph{MaxNumIter}=20). If convergence is reached
by the end of the $20^{th}$ iteration, we record the number of users
that transmit on each carrier; otherwise, we assume there is no
equilibrium. For our simulations, $P_{max}$ is assumed to be very
large which translates to having no transmit power limit for the
user terminals.

\begin{figure}
\centering
\includegraphics[width=3.5in]{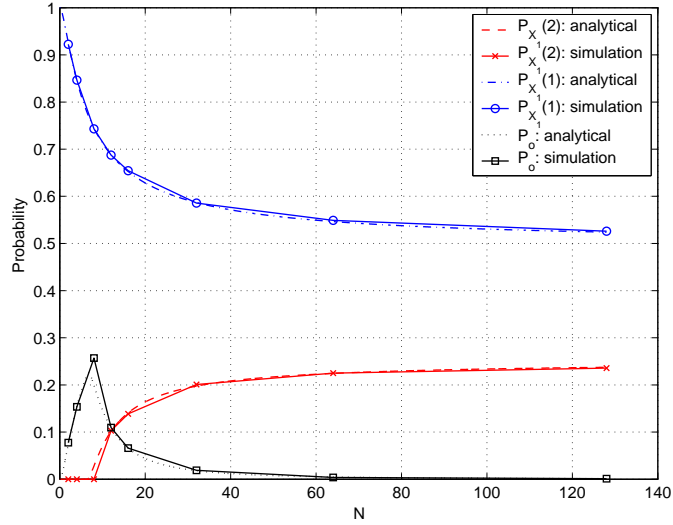}
\caption{Probability of having $m$ users on the first carrier at
equilibrium, $P_{X_1}(m)$, for $m=1,2$ as well as probability of
having no equilibrium are shown as functions of the processing gain
$N$ for the matched filter for the two-user two-carrier case (i.e.,
$D=2$ and $K=2$).}\label{probMF}
\end{figure}

\begin{figure}
\centering
\includegraphics[width=3.5in]{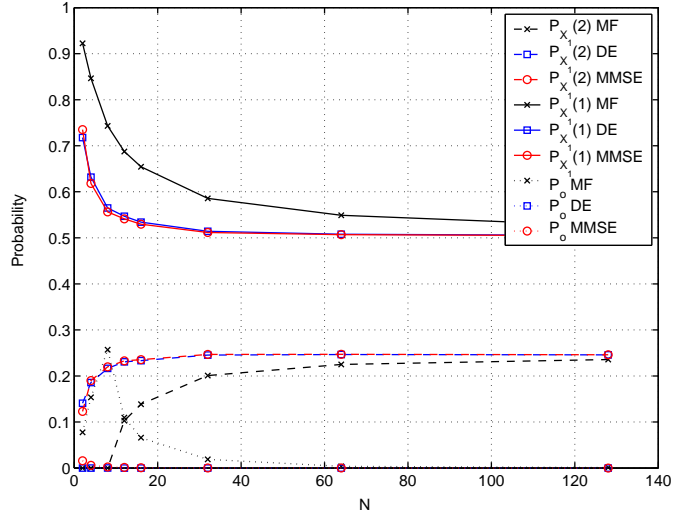}
\caption{Probability of having $m$ users on the first carrier at
equilibrium, $P_{X_1}(m)$, for $m=1,2$ as well as probability of
having no equilibrium are shown as functions of the processing gain
$N$ for the matched filter (MF), the decorrelator (DE) and the MMSE
detector for the two-user two-carrier case (i.e., $D=2$ and
$K=2$).}\label{prob-AllD2K2}
\end{figure}

It is observed that the BMP algorithm proposed in Section
\ref{algorithm} converges whenever a Nash equilibrium exists. Recall
that $P_{X_1}(m)$ represents the probability that exactly $m$ users
transmit on the first carrier at equilibrium. Fig. \ref{probMF}
shows $P_{X_1}(2)$, $P_{X_1}(1)$ and $P_o$ (probability of no
equilibrium) as a function of the processing gain $N$ for the
matched filter. The analytical expressions obtained in Section
\ref{special case} (see \eqref{eq17a}--\eqref{eq17c}) are also
plotted. We see that there is a close agreement between the
simulation results and the analytical values. It is also observed
that as $N$ becomes large, $P_o$ approaches zero. For $N=16$, for
example, the probability that a Nash equilibrium exists is about
93\%. Since $P_{X_1}(0)$ is identical to $P_{X_1}(2)$, it is not
shown in the figure. For comparison, we have shown the probabilities
for the matched filter, the decorrelator and the MMSE detector in
Fig.~\ref{prob-AllD2K2}. It should be noted that $P_o$ for the
decorrelator is always 0 (as long as $N\geq K$) and $P_o$ is almost
always zero for the MMSE detector (except for low values of $N$).
This means that for these two receivers a Nash equilibrium almost
always exists.

\begin{figure}
\centering
\includegraphics[width=3.5in]{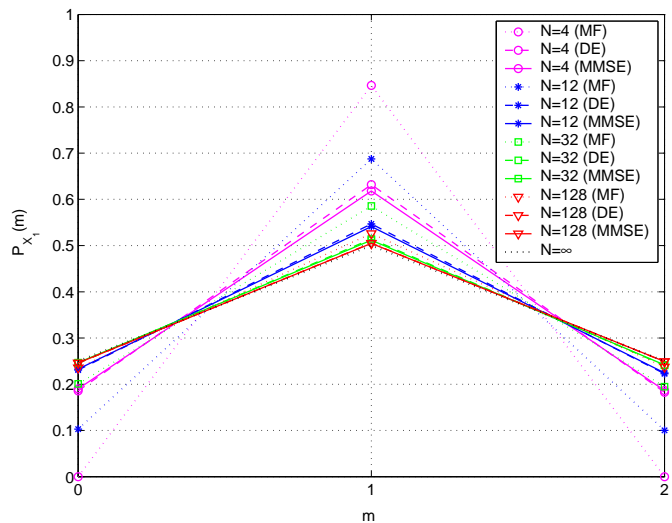}
\caption{The (pseudo) probability mass function of $X_1$ for
different values of processing gain $N$ for the two-user
two-carrier case (i.e., $D=2$ and $K=2$) for the matched filter
(MF), the decorrelator (DE), and the MMSE detector. $X_1$ is a
random variable representing the number of users transmitting on
the first carrier at Nash equilibrium.}\label{PMF-AllD2K2}
\end{figure}

\begin{figure}
\centering
\includegraphics[width=3.5in]{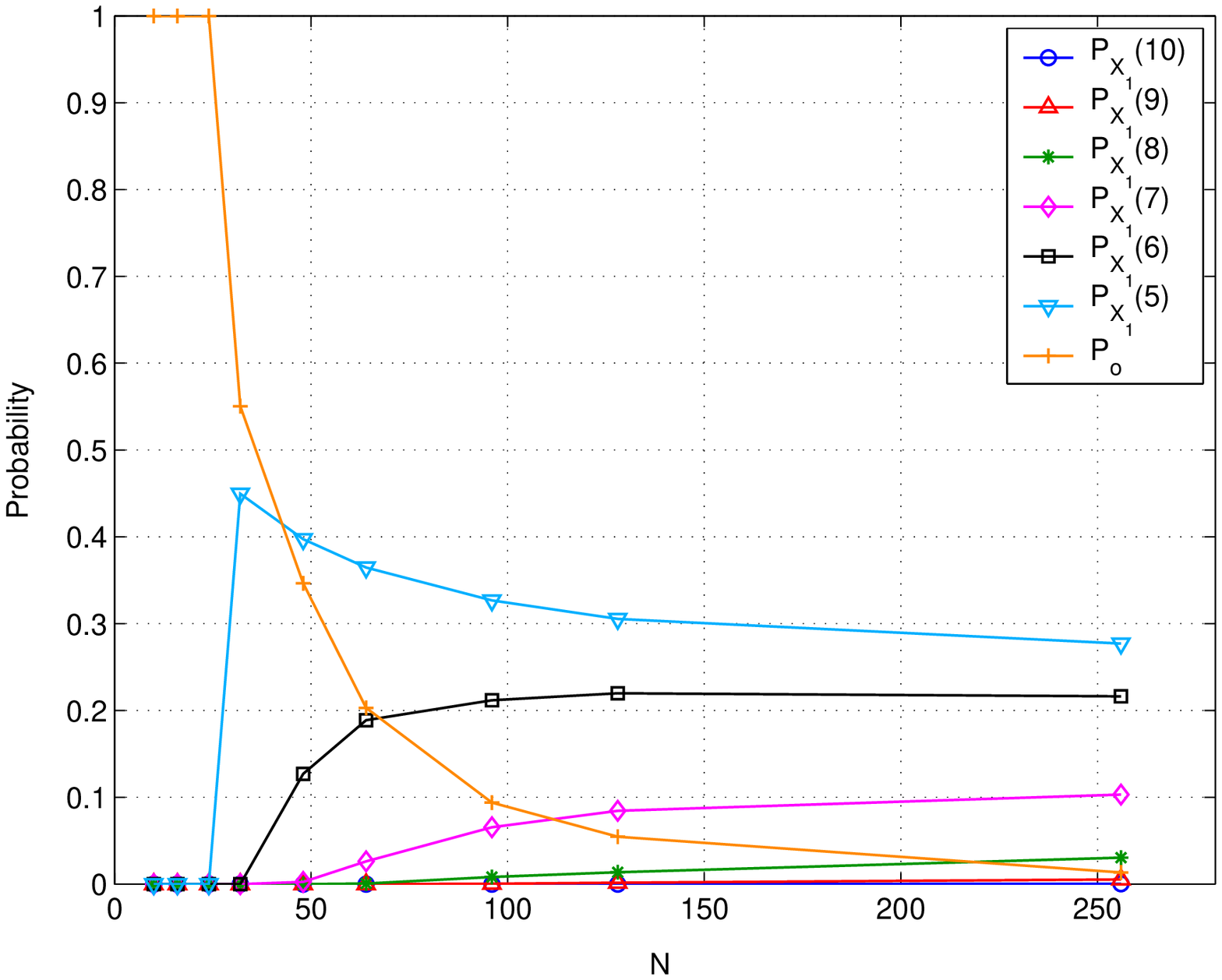}
\caption{Probability of having $m$ users on the first carrier at
equilibrium, $P_{X_1}(m)$, for $m=5, 6,\cdots, 10$ as well as
probability of having no equilibrium are shown as functions of the
processing gain $N$ for the matched filter for the ten-user
two-carrier case (i.e., $D=2$ and $K=10$).}\label{probMF-K10}
\end{figure}

Fig. \ref{PMF-AllD2K2} shows $P_{X_1}(m)$ as a function of $m$ for
different values of $N$. The results are shown for the matched
filter, the decorrelator and the MMSE receiver. We can see from
the figure that as the processing gain increases, the pseudo PMF
of $X_1$ becomes wider (i.e., it has a larger variance). This is
because for larger values of $N$, the system becomes more tolerant
towards interference. Therefore, the probability with which the
two users are able to transmit on the same carrier at equilibrium
increases. Since the decorrelating and MMSE detectors are more
effective in suppressing interference, for the same processing
gain, the corresponding PMF's are wider than that of the matched
filter. We repeat the above experiment for the case of two
carriers and 10 users. Fig.~\ref{probMF-K10} shows $P_{X_1}(m)$
for the matched filter as a function of the processing gain. Due
to symmetry, we have only plotted the probabilities for $m=5,
6,\cdots, 10$. The probability that no equilibrium exists (i.e.,
$P_o$) is also shown. Similar trends as those observed in the case
of $K=2$ are also seen here. We observe that here again as $N$
becomes large, $P_o$ approaches zero. It should be noted that when
$N$ is small, $\gamma^*$ cannot be achieved simultaneously by all
the users at the output of the matched filters. Therefore, users
keep increasing their transmit powers and hence no equilibrium is
reached. This is the case until $N$ becomes large enough so that
it can accommodate at least 5 users on each carrier (i.e.,
$N>25.6$). In practice, however, if $N$ is not large enough, some
or all users end up transmitting at the maximum power.

\begin{figure}
\centering
\includegraphics[width=3.5in]{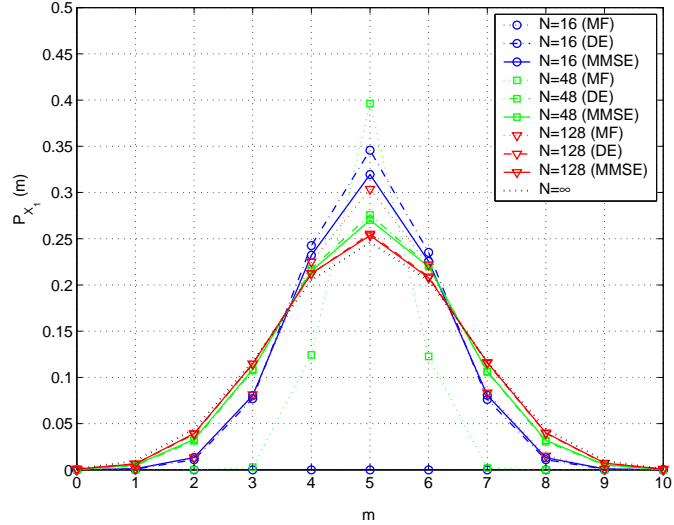}
\caption{The (pseudo) probability mass function of $X_1$ for
different values of the processing gain $N$ for the ten-user
two-carrier case (i.e., $D=2$ and $K=10$) for the matched filter
(MF), the decorrelator (DE), and the MMSE detector. $X_1$ is a
random variable representing the number of users transmitting on
the first carrier at Nash equilibrium.}\label{PMF-AllK10}
\end{figure}

\begin{figure}
\centering
\includegraphics[width=3.5in]{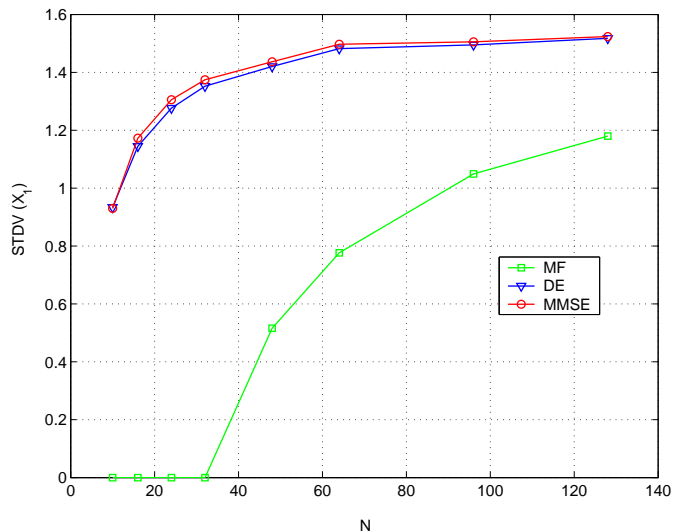}
\caption{The standard deviation for $X_1$ based on the pseudo
probability mass function as a function of the processing gain $N$
for the ten-user two carrier case (i.e., $D=2$ and $K=10$) for the
matched filter (MF), the decorrelator (DE), and the MMSE detector.
$X_1$ is a random variable representing the number of users
transmitting on the first carrier at Nash
equilibrium.}\label{stdvK10}
\end{figure}

In Fig. \ref{PMF-AllK10}, $P_{X_1}(m)$ is plotted as a function of
$m$ for the matched filter as well as the decorrelating and MMSE
detectors for different values of $N$.  The asymptotic approximation
for $P_{X_1}(m)$, given by (\ref{eq18}), is also shown. We see from
the figure that as the processing gain increases, the pseudo PMF of
$X_1$ becomes wider because the system becomes more
interference-tolerant. Also, the equilibria for which the allocation
of users to the carriers is highly asymmetric (e.g., $m=9$ and 10)
are unlikely to happen. In other words, with a high probability, the
users are evenly distributed between the two carriers. As expected,
the pseudo PMF's of the decorrelator and the MMSE detector are wider
than those of the matched filer and are more closely approximated by
\eqref{eq18}, especially as $N$ becomes large. Based on the pseudo
PMF's of $X_1$, we have plotted the standard deviation of $X_1$ for
the matched filter, the decorrelator and the MMSE detector as a
function of $N$ (see Fig. \ref{stdvK10}). It is observed that the
standard deviation of $X_1$ increases as the processing gain
increases and the values are higher for the multiuser detectors as
compared to those of the matched filter. This means that it is more
likely for the decorrelator and the MMSE detector to have a more
non-uniform distribution of users over the carriers at equilibrium.
This, of course, makes sense because the multiuser detectors are
more powerful in combating the multiple-access interference as
compared to the matched filter.

\begin{figure}
\centering
\includegraphics[width=3.5in]{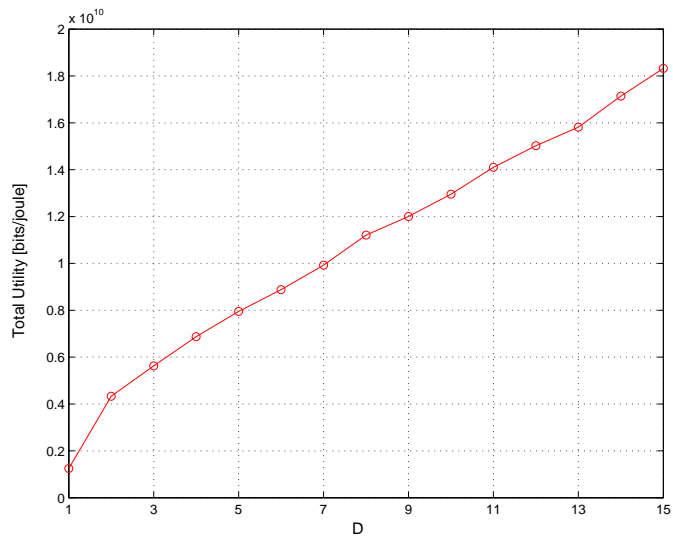}
\caption{Total utility vs. number of carriers, $D$, with 30 users
for a fixed bandwidth. The processing gain is 256 when $D=1$ and
decreasing proportionally as $D$ increases (i.e.,
$N=\frac{256}{D}).$}\label{UvsD}
\end{figure}
\begin{figure}
\centering
\includegraphics[width=3.5in]{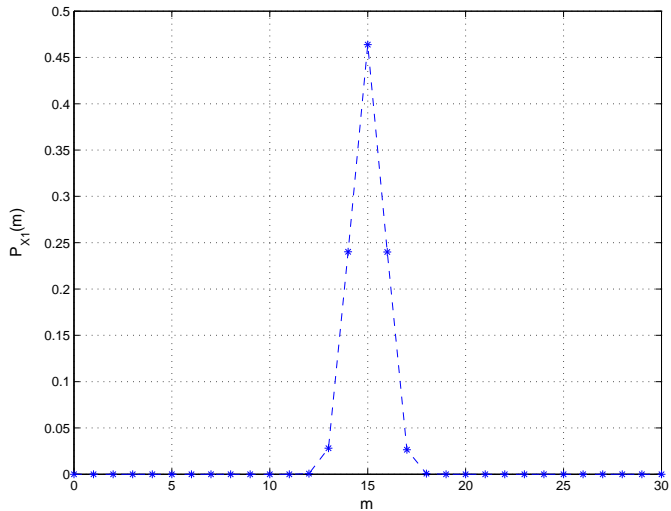}
\caption{Probability of having $m$ users on the first carrier at
Nash equilibrium, $P_{X_1}(m)$, for the thirty-user, two-carrier
case with processing gain equal to 128 (i.e., $D=2$, $K=30$ and
$N=128$).}\label{UvsD2}
\end{figure}

We now investigate the effect of the number of carriers on the
energy efficiency of the system. For a fixed bandwidth, as the
number of carriers, $D$, increases, the number of independent
channels for each user increases. Hence, the quality of the
``best" channel for each user improves. On the other hand, since
the bandwidth is fixed, the processing gain for each carrier
decreases as the number of carriers increases.  Let us consider a
system with 30 users ($K=30$). We fix the bandwidth and change the
number of carriers from 1 (i.e., single carrier) to 15. The
processing gain for the case of $D=1$ is assumed to be 256. For
each value of $D$, we run the BMP algorithm  for 20 iterations and
compute the total utility of the system at the end of the
$20^{th}$ iteration. We repeat this for 20~000 channel
realizations and average the results. Fig.~\ref{UvsD} shows the
average total utility vs. number of carriers. It is observed from
the figure that as the number of carriers increases, the total
utility also increases. We might have expected the utility to
decrease after a certain point because of the reduction in the
processing gain. However, a smaller processing gain results in a
more even distribution of users among the carrier. To demonstrate
this, Fig.~\ref{UvsD2} shows the $P_{X_1}(m)$ as a function of $m$
for two-carrier case with $N=128$. This plot is obtained by
counting the number of users on each carrier at the end of the
$20^{th}$ iteration of the algorithm. It is seen that most of the
time the users are equally distributed between the two carriers.
Similarly, we can say that with $D$ carriers, most of the time, we
will have $K/D$ users transmitting on each carrier. Therefore, the
system does not suffer from excessive interference although the
processing gain drops as $D$ increases. Since a system with a
larger number of carriers benefits more from diversity, the total
utility increases as the number of carriers increases.

We have also run simulations for the scenario in which the number of
users and the number of carriers both increase but their ratio and
the processing gain per carrier stay fixed. In particular, we
considered the cases of $D=1, 2, 4,$ and 8, with $N=64$ and $K/D=2$.
For each case, we ran the BMP algorithm for 20~000 channel
realizations with $MaxNumIter=10D$. For each channel realization, if
a stable state were reached at the end of the last iteration, we
would count the number of users on each carrier and record the
values. While the results of the simulation are difficult to show
graphically, we briefly describe some of the observed trends. We saw
from the simulation that as $D$ increased, the probability that no
Nash equilibrium existed also increased. On the other hand,
asymmetric Nash equilibria became less likely. This is because as
$D$ increases, the number of users also increases but the processing
gain stays the same. Therefore, larger transmit powers are required
to achieve $\gamma^*$ for asymmetric cases. As a result, non-uniform
Nash equilibria become less probable.

\begin{figure}
\centering
\includegraphics[width=3.5in]{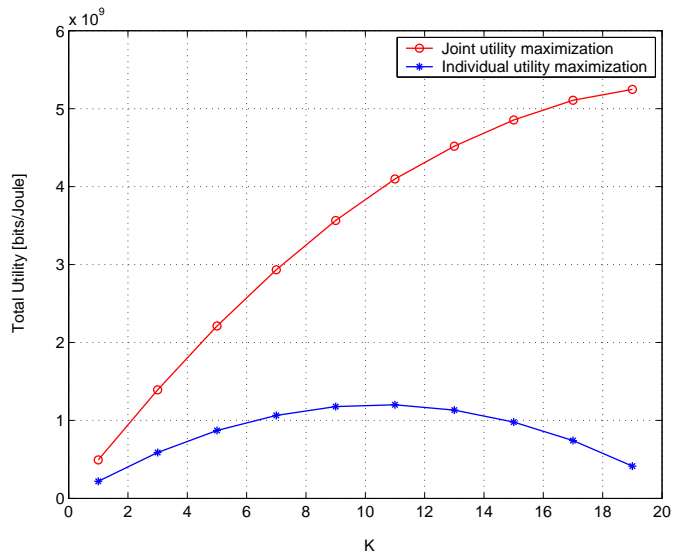}
\caption{Total utility vs. number of users, $K$, for the
two-carrier case with processing gain equal to 128 (i.e., $D=2$
and $N=128$).}\label{util-compare}
\end{figure}

We now compare our proposed approach, which jointly maximizes each
user's utility over all the carriers, with the approach that
maximizes each user's utility independently over each carrier. In
the joint maximization approach each user transmits only on the
carrier that has the best effective channel whereas in the other
case, all users transmit on all the carriers such that the output
SINR on each carrier is $\gamma^*$. We consider a system with two
carriers and $N=128$. We fix $K$ and compute the sum of the
utilities achieved by all users for 20~000 channel realizations. The
utility for each user is the ratio of the total number of
transmitted bits over the two carriers to the total energy consumed.
Fig. \ref{util-compare} shows the average total utility vs. $K$ for
the two approaches. We see a significant improvement in the achieved
utility when joint maximization over all carriers is used. This is
because when all the users transmit on every carrier, they cause
unnecessary interference. To achieve $\gamma^*$, each user is hence
forced to transmit at a higher power level which in turn results in
a considerable reduction in the overall utility. In the joint
optimization approach, each user transmits only on its ``best"
carrier. This way, the users perform a distributed interference
avoidance mechanism which results in a higher overall utility.

\section{Conclusion} \label{conclusion}
We have modeled power control for multi-carrier CDMA systems as a
non-cooperative game in which each user needs to decide how much
power to transmit over each carrier to maximize its overall utility.
The utility function has been defined as the overall throughput
divided by the total transmit power over all the carriers and has
units of bits per Joule. This game is particularly difficult to
analyze because users' strategies are vectors (rather than scalars)
and the utility function is not a quasiconcave function of the
user's strategy. For this utility function, we have shown that at a
Nash equilibrium each user transmits only on the carrier that has
the best ``effective channel" for that user. In addition, we have
derived conditions for existence and uniqueness of Nash equilibrium
and we have characterized the distribution of the users over the
carriers at equilibrium. Basically, satisfaction of a set of
inequalities by the relative channel gains of each user is
sufficient for the existence of a Nash equilibrium. We have also
proposed an iterative and distributed algorithm for reaching the
equilibrium (when it exists). We have shown that our proposed
approach results in a significant improvement in the total network
utility achieved at equilibrium compared to a single-carrier system
and also to a multi-carrier system in which users maximize their
utilities over each carrier independently.

\appendices

\section{ Proof of Convergence of the BMP algorithm for
the Two-user Two-carrier Case} \label{appendixI}

We prove that for the case of two users and two carriers, the BMP
algorithm  converges to an equilibrium whenever it exists. For the
sake of simplicity, here we assume a matched filter receiver.
However, the proof can easily be generalized to other linear
receivers. Because of symmetry, we need only to consider two of
the four possible Nash equilibria, namely, the equilibrium in
which both users transmit on the first carrier, denoted by $(12,\
\ )$, and the one in which user~1 transmits on the first carrier
and user~2 transmits on the second carrier, denoted by $(1,2)$.

\subsection{The case in which $(12,\ \ )$ is the Nash equilibrium}

In this case, we have $\frac{h_{11}}{h_{12}}>\Theta_2$ and
$\frac{h_{21}}{h_{22}}>\Theta_2$. The received powers at
equilibrium are given by $q_{11}^*=p_{11}^* h_{11}= {\gamma^*
\sigma^2}\Theta_2$ and $q_{21}^*=p_{21}^* h_{21}={\gamma^*
\sigma^2}\Theta_2$.

\subsubsection{Starting from $(12,\ \ )$} \label{app1-1}

If user~1 (user~2) starts first and the received power for user~2
(user~1) is less than ${\gamma^*\sigma^2}\Theta_2$, user~1
(user~2) stays on the first carrier with its updated received
power being less than ${\gamma^* \sigma^2}\Theta_2$ but larger
than that of user~2 (user~1). Since the received powers of both
users are less than ${\gamma^* \sigma^2}\Theta_2$, user~1 and
user~2 both continue to stay on the first carrier and update their
powers until the equilibrium is reached.

If the received powers are such that user~1 (user~2) jumps to the
second carrier, then user~2 (user~1) stays on the first carrier
with its received power equal to ${\gamma^* \sigma^2}$. Now,
because $\frac{h_{11}}{h_{12}}>\Theta_2$
($\frac{h_{21}}{h_{22}}>\Theta_2$), user~1 (user~2) jumps back to
the first carrier with $q_{11}={\gamma^* \sigma^2}/\Theta_0$
($q_{21}={\gamma^* \sigma^2}/\Theta_0$) which is less than
${\gamma^* \sigma^2}\Theta_2$. From this point, both users stay on
the first carrier and update their powers until the equilibrium is
reached.

\subsubsection{Starting from $(\ \ ,12)$}

Since $\frac{h_{11}}{h_{12}}>\Theta_2$
($\frac{h_{21}}{h_{22}}>\Theta_2$), if user~1 (user~2) starts
first, it will jump to carrier one with its received power equal
to ${\gamma^* \sigma^2}$. Now, user~2 (user~1) jumps to carrier
one with $q_{21}={\gamma^* \sigma^2}/\Theta_0$ ($q_{11}={\gamma^*
\sigma^2}/\Theta_0$) which is less than ${\gamma^*
\sigma^2}\Theta_2$. From this point, both users stay on the first
carrier and update their powers until the equilibrium is reached.

\subsubsection{Starting from $(1,2)$}\label{app1-2}

If user~1 starts first, it will stay on the first carrier with
$q_{11}={\gamma^* \sigma^2}$. Then user~2 will jump on first
carrier with $q_{21}={\gamma^* \sigma^2}/\Theta_0$. Therefore, as
before, both users stay on the first carrier and update their
powers until the equilibrium is reached.

If user~2 starts first and stays on the second carrier, then we
are back to the case that was just described. If user~2 jumps to
the first carrier and user~1 stays on the first carrier, then we
are back to the case that was already explained in
Section~\ref{app1-1}. On the other hand, if user~2 jumps to the
first carrier and user~1 jumps to the second carrier, then user~2
will stay on the first carrier with $q_{21}={\gamma^* \sigma^2}$.
After this, user~1 jumps back to the first carrier with
$q_{11}={\gamma^* \sigma^2}/\Theta_0$. Thus, again, both users
stay on the first carrier and update their powers until the
equilibrium is reached.

\subsubsection{Starting from $(2,1)$}

The argument is similar to Section~\ref{app1-2} due to symmetry.

\subsection{The case in which $(1,2)$ is the Nash equilibrium}

Let us consider the case in which $(1,2)$ is the only Nash
equilibrium. This corresponds to $\frac{h_{11}}{h_{12}}>\Theta_0$
and $\frac{h_{21}}{h_{22}}<\Theta_0$ or
$\frac{h_{11}}{h_{12}}>1/\Theta_0$  and
$\frac{h_{21}}{h_{22}}<1/\Theta_0$. For this equilibrium, we have
$q_{11}^* = q_{22}^* ={\gamma^* \sigma^2}$.

\subsubsection{Starting from $(12,\ \ )$} \label{app1-3}

If user~1 starts first and stays on the first carrier, then user~2
will jump to carrier two with $q_{22} ={\gamma^* \sigma^2}$. As a
result, user~1 will stay on the first carrier with $q_{11}^*
={\gamma^* \sigma^2}$ and hence the equilibrium is reached.

If user~1 starts first but jumps on the second carrier, then
because we are assuming that $(1,2)$ is the only Nash equilibrium
(i.e., $\frac{h_{11}}{h_{12}}>\Theta_0$ and
$\frac{h_{21}}{h_{22}}<\Theta_0$ or
$\frac{h_{11}}{h_{12}}>1/\Theta_0$  and
$\frac{h_{21}}{h_{22}}<1/\Theta_0$), user~2 will also jump on the
second carrier. As a result, user~1 will jump back to carrier one
with $q_{11}^* ={\gamma^* \sigma^2}$ and user~2 will stay on
carrier two with $q_{22}^* ={\gamma^* \sigma^2}$. Hence, the
equilibrium is reached.

If user~2 starts first, it will jump on the second carrier with
$q_{22} ={\gamma^* \sigma^2}$. As a result, user~1 will stay on
the first carrier with $q_{11}^* ={\gamma^* \sigma^2}$ and hence
equilibrium is reached.

\subsubsection{Starting from $(\ \ ,12)$}

The argument is similar to Section~\ref{app1-3} due to symmetry.

\subsubsection{Starting from $(1,2)$}

If user~1 (user~2) starts first, it will stay on the first
(second) carrier with $q_{11} ={\gamma^* \sigma^2}$ ($q_{22}
={\gamma^* \sigma^2}$). As a result, user~1 (user~2) will stay on
the first (second) carrier and hence equilibrium is reached.

\subsubsection{Starting from $(2,1)$}

If user~1 (user~2) starts first and jumps to the first (second)
carrier, user~2 (user~1) will jump to the second (first) carrier
with $q_{22} ={\gamma^* \sigma^2}$ ($q_{11} ={\gamma^*
\sigma^2}$). Thus, user~1 (user~2) will stay on the first (second)
carrier and the equilibrium is reached.

If user~1 (user~2) starts first but it stays on the second (first)
carrier, then we will have $q_{12} ={\gamma^* \sigma^2}$ ($q_{21}
={\gamma^* \sigma^2}$). But because we are assuming that $(1,2)$
is the only Nash equilibrium (i.e.,
$\frac{h_{11}}{h_{12}}>\Theta_0$ and
$\frac{h_{21}}{h_{22}}<\Theta_0$ or
$\frac{h_{11}}{h_{12}}>1/\Theta_0$  and
$\frac{h_{21}}{h_{22}}<1/\Theta_0$), user~2 (user~1) jumps to
carrier two (one) with $q_{22}^* ={\gamma^* \sigma^2}/\Theta_0$
($q_{22}^* ={\gamma^* \sigma^2}/\Theta_0$). As a result, user~1
(user~2) jumps to the first (second) carrier with $q_{11}^*
={\gamma^* \sigma^2}$ ($q_{22}^* ={\gamma^* \sigma^2}$). Thus,
user~2 (user~1) will stay on the second (first) carrier and the
equilibrium is reached.

It can be seen from the above arguments that if $(1,2)$ and
$(2,1)$ are both Nash equilibria, the algorithm converges to one
of them depending on the initial condition.

\section{Proof of Convergence of the BMP algorithm for
the Two-user $D$-carrier Case} \label{appendixII}

We prove the convergence of the BMP algorithm for the case of two
users and three carriers. Generalization to the 2-user $D$-carrier
case is straightforward. In the 2-user 3-carrier case, there are 9
possible Nash equilibria. For the sake of simplicity, we assume a
matched filter receiver. However, the proof can easily be
generalized to other linear receivers. Here, we focus on the
$(12,\ \ ,\ \ )$ Nash equilibrium. An argument similar to the one
given here and the ones used for the 2-user~2-carrier case can be
used for other equilibria.

\subsection{The case in which $(12,\ \ ,\ \ )$ is the Nash equilibrium}

In this case, we have $\frac{h_{11}}{h_{12}}>\Theta_2$,
$\frac{h_{11}}{h_{13}}>\Theta_2$, $\frac{h_{21}}{h_{22}}>\Theta_2$
and $\frac{h_{21}}{h_{23}}>\Theta_2$. The received powers at
equilibrium are given by $q_{11}^*=p_{11}^* h_{11}= {\gamma^*
\sigma^2}\Theta_2$ and $q_{21}^*=p_{21}^* h_{21}={\gamma^*
\sigma^2}\Theta_2$.

\subsubsection{Starting from $(12,\ \ ,\ \ )$} \label{app2-1}

If user~1 (user~2) starts first and the received power for user~2
(user~1) is less than ${\gamma^*\sigma^2}\Theta_2$, user~1
(user~2) stays on the first carrier with its updated received
power being less than ${\gamma^* \sigma^2}\Theta_2$ but larger
than that of user~1 (user~2). Since the received powers of both
users are less than ${\gamma^* \sigma^2}\Theta_2$, user~1 and
user~2 both continue to stay on the first carrier and update their
powers until the equilibrium is reached.

If the received powers are such that user~1 (user~2) jumps to the
second or third carrier, then user~2 (user~1) stays on the first
carrier with its received power equal to ${\gamma^* \sigma^2}$.
Now, because $\frac{h_{11}}{h_{12}}>\Theta_2$,
$\frac{h_{11}}{h_{13}}>\Theta_2$, $\frac{h_{21}}{h_{22}}>\Theta_2$
and $\frac{h_{21}}{h_{23}}>\Theta_2$, user~1 (user~2) jumps back
to the first carrier with $q_{11}={\gamma^* \sigma^2}/\Theta_0$
($q_{21}={\gamma^* \sigma^2}/\Theta_0$) which is less than
${\gamma^* \sigma^2}\Theta_2$. From this point, both users stay on
the first carrier and update their powers until the equilibrium is
reached.

\subsubsection{Starting from $(\ ,12,\ )$} \label{app2-2}

Since $\frac{h_{11}}{h_{12}}>\Theta_2$
($\frac{h_{21}}{h_{22}}>\Theta_2$), if user~1 (user~2) starts
first, it will jump to carrier one with its received power equal
to ${\gamma^* \sigma^2}$. Now, user~2 (user~1) jumps to carrier
one with $q_{21}={\gamma^* \sigma^2}/\Theta_0$ ($q_{11}={\gamma^*
\sigma^2}/\Theta_0$) which is less than ${\gamma^*
\sigma^2}\Theta_2$. From this point, both users stay on the first
carrier and update their powers until the equilibrium is reached.

\subsubsection{Starting from $(\ ,\ ,12)$}

The argument is similar to Section~\ref{app2-2} due to symmetry.

\subsubsection{Starting from $(1,2,\ \ )$}\label{app2-3}

If user~1 starts first, it will stay on the first carrier with
$q_{11}={\gamma^* \sigma^2}$. Then user~2 will jump on first
carrier with $q_{21}={\gamma^* \sigma^2}/\Theta_0$. Therefore, as
before, both users stay on the first carrier and update their
powers until the equilibrium is reached.

If user~2 starts first and stays on the second carrier, then we
are back to the case that was just described. If user~2 jumps to
the third carrier, user~1 stays on the first carrier with
$q_{11}={\gamma^* \sigma^2}$. Then user~2 jumps to the first
carrier and both users stay there until equilibrium is reached.

If user~2 jumps to the first carrier and user~1 stays on the first
carrier, then we are back to the case that was already explained
in Section~\ref{app2-1}. On the other hand, if user~2 jumps to the
first carrier and user~1 jumps to the second  or third carrier,
then user~2 will stay on the first carrier with $q_{21}={\gamma^*
\sigma^2}$. After this, user~1 jumps back to the first carrier
with $q_{11}={\gamma^* \sigma^2}/\Theta_0$. Thus, again, both
users stay on the first carrier and update their powers until the
equilibrium is reached.

\subsubsection{Starting from $(1,\ ,2)$}

The argument is similar to Section~\ref{app2-3} due to symmetry.

\subsubsection{Starting from $(2,1,\ \ )$}

The argument is similar to Section~\ref{app2-3} due to symmetry.

\subsubsection{Starting from $(2,\ ,1)$}

The argument is similar to Section~\ref{app2-3} due to symmetry.

\subsubsection{Starting from $(\ ,1,2)$}\label{app2-4}

If user~1 (user~2) starts first, it will jump to carrier one with
$q_{11}={\gamma^* \sigma^2}$ ($q_{21}={\gamma^* \sigma^2}$). Then,
user~2 (user~1) jumps to the first carrier. Both user will then
stay on the first carrier until equilibrium is reached.

\subsubsection{Starting from $(\ ,2,1)$}

The argument is similar to Section~\ref{app2-4} due to symmetry.

\section{Proof of Convergence of the BMP algorithm for the
Three-user two-carrier Case} \label{appendixIII}

We prove the convergence of the BMP algorithm for the case of
three users and two carriers. In this case, there are 27 possible
Nash equilibria. Here, we focus on the $(123,\ \ )$ and $(12,3)$
Nash equilibria. Similar arguments can be used for other
equilibria. \vspace{-0.4cm}

\subsection{The case in which $(123,\ \ )$ is the Nash equilibrium}

In this case, we have $\frac{h_{11}}{h_{12}}>\Theta_3$,
$\frac{h_{21}}{h_{22}}>\Theta_3$ and
$\frac{h_{31}}{h_{32}}>\Theta_3$. The received powers at
equilibrium are given by $q_{11}^*=q_{21}^*=q_{31}^*={\gamma^*
\sigma^2}\Theta_3$. Here, we only discuss the convergence for two
possible starting points. Similar arguments can be applied to the
other cases as well.

\subsubsection{Starting from $(123,\ \ )$} \label{app3-1}

If the powers are such that the three users stay on the first
carrier, then they keep updating their powers until they reach the
equilibrium.

On the other hand, the powers can be such that one of the users
jumps to the second carrier. Without loss of generality, let us
assume that user~3 jumps to the second carrier, i.e., let us
assume that we have reached (12,3) and user~1 has to now update
its power.
\begin{itemize}
  \item Now, if user~1 also jumps to the second carrier, user~2 stays on the
first carrier with $q_{21}=\gamma^* \sigma^2$. As a result, user~3
jumps back to carrier one with $q_{31}=\gamma^*
\sigma^2/\Theta_0$. Thus, user~1 also jumps to the first carrier
and the users stay there until the equilibrium is reached.
  \item Now let us assume that user~3 jumps but user~1 stays.
\begin{itemize}
  \item If $q_{21}\leq \gamma^* \sigma^2 \Theta_3$, then after user~1
updates its power, we will have $q_{11} \leq \gamma^* \sigma^2
\Theta_3$. Therefore, user~1 and 2 both stay on the first carrier
and user~3 also jumps back to carrier one. The three users stay on
this carrier until equilibrium is reached.
   \item  If $q_{21}>\gamma^* \sigma^2\Theta_3$, then after user~1 updates its
power, we will have $q_{11}<q_{12}$. If user~2 stays, then user~1
will also stay. In this case, the powers of users 1 and 2 decrease
until user~3 jumps back to the first carrier. In this case,
$q_{31}\leq \max\{q_{11}, q_{21}\}$. One or two users may still
jump to the second carrier but the powers of the users on the
first carrier keep decreasing. As a result, the users will
eventually come back to the first carrier and at some point, the
powers become such that all three users stay on the first carrier
until the equilibrium is reached. On the other hand, if user~2
jumps to the second carrier and user~3 stays on the second
carrier, then user~1 stays on the first carrier with $q_{11}\leq
\gamma^* \sigma^2$. As a result, users 2 and 3 jump back to the
first carrier and all three users stay there until an equilibrium
is reached. In the case where user~2 jumps on the second carrier
but user~3 jumps back to the first carrier, we will have
$q_{31}<q_{11}$. One or two may still continue to jump to the
second carrier but the powers of the users on the first carrier
keep decreasing. Eventually, a point is reached where all three
users come back to the first carrier and stay there until the
equilibrium is reached.
\end{itemize}

\end{itemize}

\subsubsection{Starting from $(1,23)$}

If user~1 starts first, it will stay on the first carrier. As a
result, users 2 and 3 will also jump to the first carrier and all
three users stay there until the equilibrium is reached.

Now, consider the case where user~2 starts first.
\begin{itemize}
  \item If user~2 jumps on the first carrier and user~3 also jumps then we are back to
the case that was discussed in Section~\ref{app3-1}. If user~2
jumps but user~3 stays on second carrier, then we are back to the
(12,3) case discussed as part of Section~\ref{app3-1}.
  \item If user~2 stays on the second carrier and user~3 also stays,
then user~1 will stay on the first carrier with $q_{11}= \gamma^*
\sigma^2$. As a result, users 2 and 3 will jump to the first
carrier and all three users stay there until an equilibrium is
reached. On the other hand, if user~2 stays on the second carrier
but user~3 jumps to the first carrier, then we will have the
(13,2) case. Although the users may still take turn in jumping to
the second carrier, the powers of the users on the first carrier
keep decreasing. Eventually, a point is reached where all three
users come back to the first carrier and stay there until the
equilibrium is reached. This is similar to the (12,3) scenario
discussed as part of Section~\ref{app3-1}.
\end{itemize}

Let us now consider the case in which user~3 starts first.
\begin{itemize}
  \item If user~3 stays on the second carrier, then user~1
will stay on the first carrier with $q_{11} = \gamma^* \sigma^2$.
As a result, users 2 and 3 will jump to the first carrier and all
three users stay there until the equilibrium is reached.
  \item If user~3 jumps to the first carrier, then we will be in the (13,2) case
 which is similar in nature to the he (12,3) scenario discussed as part of
Section~\ref{app3-1}.
\end{itemize}

Similar arguments can be used for the remaining 25 starting
points. It should be noted that symmetry can be used to reduce the
number of cases that need to be considered.\vspace{-0.1cm}

\subsection{The case in which $(12,3)$ is the Nash equilibrium}

Let us consider the case of
$\frac{h_{11}}{h_{12}}>\Theta_2/\Theta_0$,
$\frac{h_{21}}{h_{22}}>\Theta_2/\Theta_0$ and
$\frac{h_{31}}{h_{32}}<\Theta_2/\Theta_0$ for which (12,3) is the
only equilibrium. Other cases can be treated similarly. For the
sake of brevity, we only discuss the (3,12) starting point.
Similar arguments can be used for other starting points.

\subsubsection{Starting from $(3,12)$}

Let us now consider the case in which user~1 starts first.
\begin{itemize}
  \item If user~1 jumps to the first carrier and user~2 also jumps,
then user~3 will jump to the second carrier with $q_{32}= \gamma^*
\sigma^2$. Now, if either user~1 or user~2 jumps to the second
carrier, user~3 still stays in the second carrier because
$\frac{h_{31}}{h_{32}}<\Theta_2/\Theta_0$. Therefore, users 1 and
2 will eventually go back to the first carrier and stay there
until the equilibrium is reached. On the other hand, if user~1
jumps to the first carrier but user~2 stays on the second carrier,
then user~3 will jump to the second carrier with $q_{32}=\gamma^*
\sigma^2/\Theta_0$. As a result, user~1 stays on the first carrier
with $q_{11}= \gamma^* \sigma^2$ and user~2 also jumps to the
first carrier. They stay there until the equilibrium is reached.

\item Now consider the case where user~1 stays on the second carrier and user~2
also stays. \begin{itemize}
              \item If user~3 jumps to the second carrier, then
users 1 and 2 jump to the first carrier and user 3 will stay on
the second carrier. The users stay there until the equilibrium is
reached.
              \item If user~3 stays on the first carrier, then we
will have $q_{31}= \gamma^* \sigma^2$. As a result, users 1 and 2
jump to the first carrier and then user~3 jumps to the second
carrier. The users stay there until the equilibrium is reached.
            \end{itemize}
\end{itemize}

Let us now consider the case in which user~2 starts first.
\begin{itemize}
  \item If user~2 stays on the second carrier and user~3 stays on
the first carrier, then user~1 jumps on the first carrier with
$q_{11}= \gamma^* \sigma^2/\Theta_0$. As a result, user~2 also
jumps to the first carrier and consequently user~3 jumps to the
second carrier. The users stay there until the equilibrium is
reached. On the other hand, if user~2 stays on the second carrier
and user~3 jumps to the second carrier, then users 1 and 2 will
jump to the first carrier. As a result, user~3 stays on the second
carrier. Users 1 and 2 will stay on the first carrier until the
equilibrium is reached.

  \item Now assume that user~2 jumps to the first carrier.
            \begin{itemize}
              \item If user~3 stays on the first carrier and user~1
jumps to the first carrier, then independent of whether user~2
jumps to the second carrier, user~3 will jump to the second
carrier. Because of this, users 1 and 2 will eventually jump to
the first carrier and stay there until the equilibrium is reached.
On the other hand, if user~3 stays on the first carrier and user~1
stays on the second carrier, then by going through the two
possibilities for user~2, it can be shown that user~3 will
eventually jump to the second carrier and users 1 and 2 will jump
to the first carrier. The users then stay there until the
equilibrium is reached.

              \item If user~3 jumps to the second carrier and user~1
stays on the second carrier, then user~2 will stay on the first
carrier with $q_{21}=\gamma^* \sigma^2$. It can be shown by going
through the two possibilities for user~3, that users 1 and 2 will
eventually end up on the first carrier and user~3 will end up on
the second carrier. The users stay there until the equilibrium is
reached.
            \end{itemize}
\end{itemize}

Let us now consider the case in which user~3 starts first. If
user~3 stays on the first carrier, then users 1 and 2 will also
jump to the first carrier. As a result, user~3 will jump to the
second carrier. The users stay there until the equilibrium is
reached. On the other hand, if user~3 jumps to the second carrier,
then users 1 and 2 will jump to the first carrier and user~3 will
then stay on the second carrier. The users stay there until the
equilibrium is reached.

The convergence for other cases can be verified using similar
arguments.


\end{document}